\documentclass[twocolumn]{aastex631}
\usepackage{amsmath}
\usepackage{amssymb}
\usepackage{graphicx}
\usepackage{epsfig}

\newcommand{\rsun}{R$_\odot$}
\newcommand{\be}{\begin{equation}}
\newcommand{\ee}{\end{equation}}

\newcommand{\domdr}{\partial\log\Omega/\partial\log r}

\shorttitle{The near-surface shear layer}
\shortauthors{Antia \& Basu}

\begin{document}

\title{Changes in the near-surface shear layer of the Sun}

\correspondingauthor{Sarbani Basu}
\email{sarbani.basu@yale.edu}

\author[0000-0001-7549-9684]{H.~M. Antia}
\affiliation{UM-DAE Centre for Excellence in Basic Sciences, University of Mumbai, Mumbai 400098, India}

\author[0000-0002-6163-3472]{Sarbani Basu}
\affiliation{Department of Astronomy,
Yale University, PO Box 208101,\\
New Haven, CT 06520-8101,USA}

\begin{abstract}

We use helioseismic data obtained over two solar cycles to determine whether there are changes in the near-surface shear layer (NSSL). We examine this by determining the radial gradient of the solar rotation rate. The radial gradient itself shows a solar-cycle dependence, and the changes are more pronounced in the active latitudes than at adjoining higher latitudes; results at the highest latitudes ($\ga 70^\circ$) are unreliable. The pattern changes with depth, even within the NSSL. We find that the near-surface shear layer is deeper at lower latitudes than at high latitudes and that the extent of the layer also shows a small solar-cycle related change.

\end{abstract}

\keywords{The Sun (1693) --- solar cycle (1487) --- solar oscillations (1515) --- helioseismology (709) --- solar rotation (1524)}

\section{Introduction} \label{sec:intro}

Helioseismic inversions for solar rotation reveal two distinct shear layers --- a very prominent one that is near the base of the convection zone, and one close to the surface \citep{schou1998}. The former, named the ``tachocline'' \citep{tachocline} is the main seat of the solar dynamo \citep[see][and references therein]{gilman2005, paulchar2010} in many solar dynamo models. The latter, the near-surface shear layer \citep[NSSL;][]{mjt1996}, is still somewhat of a mystery.  The first evidence for the NSSL came from the observation that emerging active regions rotate faster than the surrounding photosphere \citep{foukal}, helioseismic analyses have since confirmed the presence of the {NSSL, and it has been a subject of many helioseismic analyses,  whether implicitly in studies of solar zonal flows \citep[e.g.,][etc.]{hmasb2001,svv2002, sbhma2003, rh2006, grad2008, rh2018, sbhma2019} or in focused studies of this layer \citep[e.g.,][etc.]{mjt, hmasb2010, barekat1, barekat2}.} 

It has been argued \citep{brandenburg, pipin} 
that the NSSL plays a role in a distributed dynamo. \citet{brandenburg} also argued that the radial-shear in the NSSL may explain the migration of the solar-activity belt towards the equator as the solar cycle progresses.  Increasingly, the role that this layer can play in solar dynamos is being studied \citep[e.g.,][etc.]{dikpati2002, mason2002, kapyla, karak2016,paradkar,jha21}. 

\citet{mjt} studied the gradient of the rotation rate in the NSSL  and found it to be nearly constant with latitude {until about $30^\circ$}; this was subsequently confirmed by \citet{barekat1} who found that the gradient is constant to much higher latitudes, up to about $60^\circ$. Both these studies only used f-mode data, limiting their results to very shallow depths. \citet{grad2008} studied changes in the gradients of the rotation rate throughout the convection zone; while they did not explicitly comment on the NSSL, they showed that at 0.95\rsun\ the radial gradient of the rotation rate shows a pattern of migrating band as a function of latitude and time that is similar to that of the zonal flows. \citet{hmasb2010} examined the radial extent of the NSSL and found some solar cycle-dependent changes, though this study used data covering only one solar cycle. They did not examine any changes to the rotation-rate gradient. \citet{barekat2} examined the rotation rate gradient as a function of time and found a well-defined time difference, again with a banded pattern migrating towards the equator with time. They found that sunspots are concentrated at latitudes where the gradient is lower than the average gradient. They claimed that this result contradicts what \citet{grad2008} found, i.e., sunspots are concentrated at latitudes where the gradient is higher than average; the \citet{barekat2} and \citet{grad2008} results, however, were at different depths inside the Sun, indicating that the change of the rotation-rate gradient is not constant over the NSSL. The NSSL extends up to about $0.95R_\odot$, and thus a time-variation that depends on radius is quite possible. Even the zonal flow pattern at low latitudes show a strong variation in phase around $0.95R_\odot$ \citep{svv2002,grad2008}.
\citet{grad2008} compared the variations of the gradient with sunspot locations at 0.95\rsun; the results of \citet{barekat2}, based only on f-mode data, are for a radius of around $0.99R_\odot$.

In this paper, we analyze helioseismic data obtained over two solar cycles to investigate changes in the NSSL as a function of time at different latitudes. Unlike \citet{mjt}, \citet{barekat1, barekat2}, we do not limit ourselves to using only the f-modes; this allows us to determine changes throughout the NSSL and also determine whether the extent of the shear-layer changes with the solar cycle.

The rest of the paper is organized as follows. We describe the data used in Section~\ref{sec:data}, the analysis technique is described in Section~\ref{sec:ana}, we present our
results in Section~\ref{sec:res}, and we discuss the implications of our results in Section~\ref{sec:disc}.

\section{Data used}
\label{sec:data}

For this work, we use solar oscillation frequencies obtained by the ground-based  Global Oscillation Network Group (GONG) \citep{gong} and the space-based Michelson Doppler Imager (MDI) on board the Solar and Heliospheric Observatory spacecraft \citep{mdi} and the Helioseismic and Magnetic Imager (HMI) \citep{hmi}  on board the Solar Dynamics Observatory. 

The data we use from the GONG project cover a period from 1995.05.05 to  2021.01.25. The data are designated by GONG ``months'', each ``month'' being 36 days long. Solar oscillation frequencies and splittings of sets starting GONG Month~2 are obtained using 108-day (i.e., 3 GONG months) time series. There is an overlap of 72 days between different data sets, i.e., GONG Month~2 frequencies were obtained from data of GONG Months~1, 2 and 3, those for GONG month~3 from GONG Months~2, 3 and 4, etc. We use data for GONG Months~2 to 260. These data are available from the GONG web-site\footnote{https://gong.nso.edu}. 
GONG datasets are restricted to modes with $l \le 150$. It should be noted that very few modes covered by the GONG data have lower turning point above $0.98R_\odot$ and hence it is not possible to infer the rotation rate above this layer using these data sets.

Data from  MDI cover the period from 1996.05.01 to 2011.04.24. Solar oscillation frequencies and splittings for these data were obtained from 72-day time series, and the sets have no overlap in time. HMI started obtaining data on 2010.04.30 and these are also obtained from 72-day time series. {MDI and HMI sets have p-mode data for $l \le 200$ and f-mode data from  $l \approx 120$ to $l=300$}. These data are available to all researchers through the Joint Science Operations Center of the Solar Dynamic Observatory\footnote{https://jsoc.stanford.edu}. We have used  MDI data from series  \texttt{mdi.vw$\_$V$\_$sht$\_$modes}, and HMI series \texttt{hmi.V$\_$sht$\_$modes}. We combine the MDI and HMI datasets to cover both solar cycles 23 and 24. There is an overlap of 1 year between the MDI and HMI data; we use HMI data for this period. The last HMI dataset that we have used is set 10288, which has an end date of 2021.05.13. 

We compare the helioseismic results with the Sunspot butterfly diagram. For this purpose, we use data on the date of emergence and latitude of the emergence of active regions from the National Oceanic and Atmospheric Administration's Solar Region Summary.

\section{Analysis technique}
\label{sec:ana}

We invert each of the data sets to determine the rotation rate as a function of depth and latitude. We adapt the 2D Regularized Least Squares (RLS) technique described by \citet{rot} to perform the inversions. The rotation rate is represented as a product of cubic B-splines in radius and $\cos(\vartheta)$, where $\vartheta$ is the colatitude. Symmetry between the two hemispheres is ensured through a boundary condition at the equator. We use 20 knots uniformly spaced in $\cos(\vartheta)$ and 50 knots uniformly spaced in the acoustic radius. The smoothing parameters were chosen to be $\lambda_r=0.0025$ and $\lambda_\theta=0.04$. Iterative refinement of the solution was done as described by \citet{antia96}, and a fixed number of 9 iterations were performed.

For the GONG data we use the first 8 odd-order splitting coefficients, $c_1,c_3,\ldots,c_{15}$, while for MDI and HMI data we use 18 splitting coefficients, $c_1,c_3,\ldots,c_{35}$. We determine the dimensionless radial gradient of the rotation rate $\partial\log\Omega/\partial\log r$ directly from the B-splines; the use of cubic B-splines obviates the need to calculate numerical derivatives. This is different from the approach of \citet{grad2008} and \citet{hmasb2010}, where the derivative was calculated by numerical differentiation of the rotation rate. Analytic derivatives reduce the noise in radial-gradient estimate and allows us to study the changes in the gradient in the NSSL.

We determine temporal changes in the rotation-rate gradient by subtracting out the average gradient for each solar cycle, i.e.,
\be
\begin{split}
\delta\left(\frac{\partial\log\Omega(r,\theta,t)}{\partial\log r}\right)= &\left(\frac{\partial\log\Omega(r,\theta,t)}{\partial\log r}\right)\\
&-\left\langle \left(\frac{\partial\log\Omega(r,\theta,t)}{\partial\log r}\right)\right\rangle,
\end{split}
\label{eq:resid}
\ee
where, $r$, $\theta$ and $t$ are respectively radius, latitude and time, $\Omega$ is the rotation rate, and the term in the angular brackets represents the time average over a solar cycle. This residual shows large values at high latitudes and to balance out the variation, we multiply the residual by $\cos\theta$ in all figures. Note that unlike \citet{grad2008}, \citet{hmasb2010}, and \citet{sbhma2019}, we use the dimensionless gradient for this work. This allows us to compare our results easily with those of \citet{mjt} and \citet{barekat1,barekat2}. Guided by our earlier work on zonal flows \citep{hmasb2013}, we average each solar cycle separately and subtract that from data sets of that cycle. Since we have very little data for solar Cycle~25, we subtract the Cycle~24 average from those sets. To smooth out fluctuations in the time series at each $r,\theta$, we apply running mean over $\pm6$ months from the central point in time.

We define the extent of the near-surface shear layer as the region where the magnitude of the dimensionless gradient is greater than 20\% of its maximum value at each epoch. The maximum is usually at the solar surface except at high latitudes; at high latitudes, the maximum gradient sometimes occurs a little below the surface. This allows us to define the position of the base of the shear layer, $r_s(\theta)$. We have tried different definitions for the extent of NSSL. A smaller value, e.g., 10\% causes problems at some epochs at the higher latitudes --- the value of $r_s$ for those epochs and latitudes suddenly becomes small (i.e., the shear-layer suddenly deepens), perhaps because of noise and also because sometimes the sign of gradient at high latitude does not change until we reach the tachocline region. Since such a sudden change is not expected, the limit of 10\% is clearly unsuitable. The 20\% value also allows us to be consistent with our previous work  \citep{hmasb2010}, except that here we use the dimensionless gradient. The use of a higher limit, e.g., 30\% doesn't lead to any significant difference, except that the average depth of the layer is reduced. It could be argued that the temporal variation in the extent of NSSL is caused by variation of the maximum value. To discount this possibility, we also considered one case where a fixed maximum magnitude of 1 was used to define the extent of the NSSL. Even in this case, the observed temporal variation is similar. Consequently, we retain the definition of \citet{hmasb2010}. We also study the temporal variations of $r_s(\theta)$ by subtracting an average over each solar cycle, just like that for the gradient.

\section{Results}
\label{sec:res}

\subsection{The radial rotation-rate gradient in the NSSL}
\label{subsec:grad}

\begin{figure}
\epsscale{1.1}
\plotone{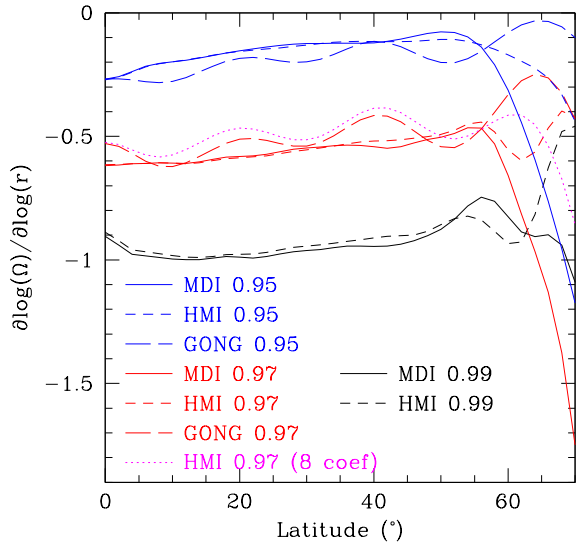}
\caption{The average gradient in the near-surface shear layer for MDI, HMI and GONG sets at different radii plotted as a function of latitude. The GONG data are averaged over Cycle~23; the results are not significantly different for Cycle~24. We also show HMI results obtained with datasets restricted to $l \le 150$ modes and 8 splitting coefficients.
Note that the absolute value of the gradient decreases deeper inside the Sun, and that there are significant differences between the GONG, MDI and HMI results at high latitudes.
}
\label{fig:ave}
\end{figure}

The time-averaged gradient of the rotation rate in the NSSL is shown in Fig.~\ref{fig:ave}, 
since GONG data sets have only a few modes with lower turning points above $0.98R_\odot$, we do not show GONG results at $0.99R_\odot$.
Note that the magnitude of the gradient decreases as one goes deeper inside the 
Sun. Like  \citet{barekat1}, we find that the gradient is almost constant up to nearly $60^\circ$ latitude, and deviate from the constant at higher latitudes. 
Our results show a steepening of the gradient with an increase in latitude; 
it is possible that this is because we do not restrict ourselves to using f-mode data and use p-mode data too. 
We also find that the MDI and HMI results do not agree at high latitudes implying that there may not be much significance of the deviation of the gradient from $-1$ at these latitudes.
The GONG results show some oscillations in the gradient as a function of the latitude. These are due to the fact that GONG data include only 8 odd-order splitting coefficients and are restricted to modes with $l\le150$. To confirm this, we repeated the analysis for HMI data using the same restriction on modes and splitting coefficients and find that they also show similar oscillations. These results are also included in the figure. These 
oscillations are most likely due to the fact that the polynomials in $\cos\vartheta$ corresponding to the first 8 splitting coefficients are not sufficient to represent the latitudinal variation in the gradient.

\begin{figure*}
\epsscale{0.85}
\plotone{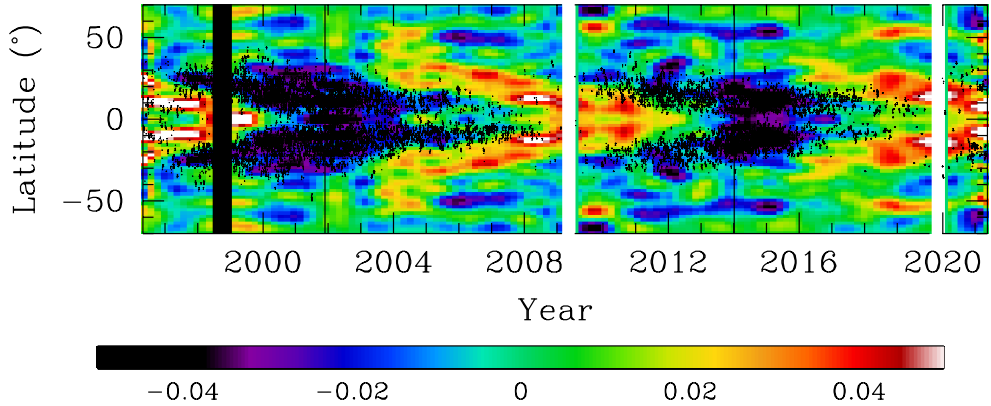}
\caption{The change in the dimensionless radial gradient of the rotation rate, $\domdr$ at 0.99\rsun\ plotted as a function of time and latitude. 
Only MDI and HMI results are shown because of the lack of resolution in GONG results at this depth.
The plot shows the residuals obtained after the time average of the gradient at each latitude is subtracted out; only Cycle~23 average has been subtracted from Cycle~23 data, Cycle~24 average has been subtracted from Cycles~24 and 25. The residuals are multiplied by $\cos\theta$ in order to avoid saturating the color scale at high latitude. The positions of sunspots are marked with black dots. The white vertical bands demarcate the solar cycles, the black vertical lines mark the maximum in each cycle and the wide vertical black band marks the time when SoHO was out of contact.  
}
\label{fig:zon99}
\end{figure*}

\begin{figure*}
\epsscale{0.85}
\plotone{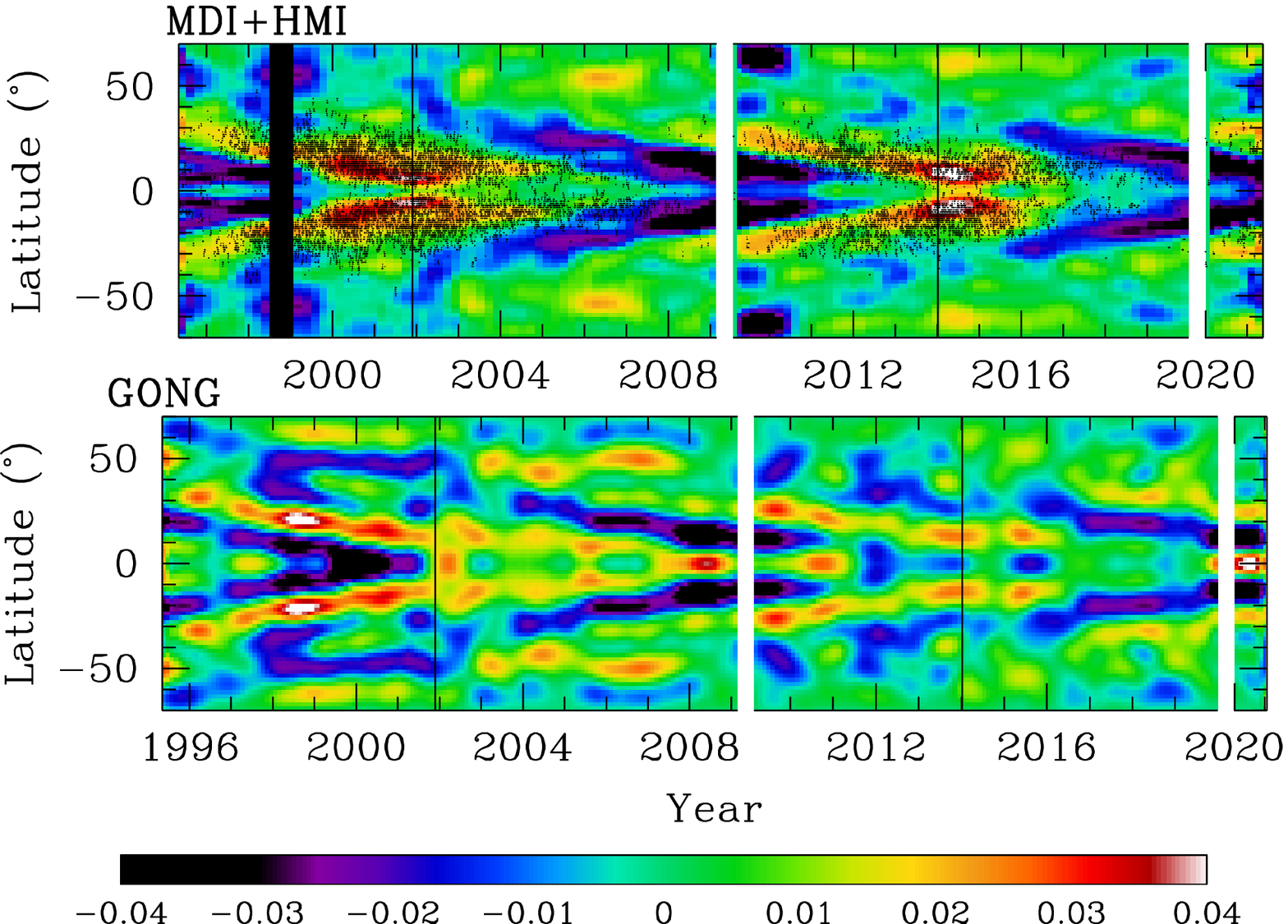}
\caption{The same as Fig.~\ref{fig:zon99}, but at 0.95\rsun. We show both MDI and HMI results (upper panel) and GONG results (lower panel). 
}
\label{fig:zon95}
\end{figure*}

Just because the average gradient is almost constant with latitude does not mean that the time-variation of the gradient is the same at all latitudes. The change in the gradient (multiplied by $\cos\theta$) at 0.99\rsun\ is shown in Fig.~\ref{fig:zon99}; this is the radius that corresponds closest to the depth of the \citet{barekat2} results making comparisons easier. 
As can be seen in the figure, the gradient in the NSSL shows a similar banded pattern as the zonal flows \citep[etc.]{rh2006, grad2008, sbhma2019, rh2020}. We also see that the sunspots are confined to regions where the gradient is lower than average. This agrees with the result of \citet{barekat2}. 

\citet{barekat2} had remarked on the disagreement between their results and the results of \citet{grad2008} with regard to the pattern of concentration of sunspots with respect to that of the radial gradient. The apparent disagreement is most likely because the results were obtained at different depths, and the phase of the pattern has a steep variation with depth in this range.
We can see from Fig.~\ref{fig:zon95}, where we show our current results at $r=0.95R_\odot$, that
 the pattern is reversed compared with that at 0.99\rsun. At this radius, sunspots coincide with the positive band, as found by \citet{grad2008} and \citet{sbhma2019}. This implies that the phase of temporal variation of the rotation gradient has a radial dependence even within the thin NSSL.

\begin{figure}
\epsscale{1.1}
\plotone{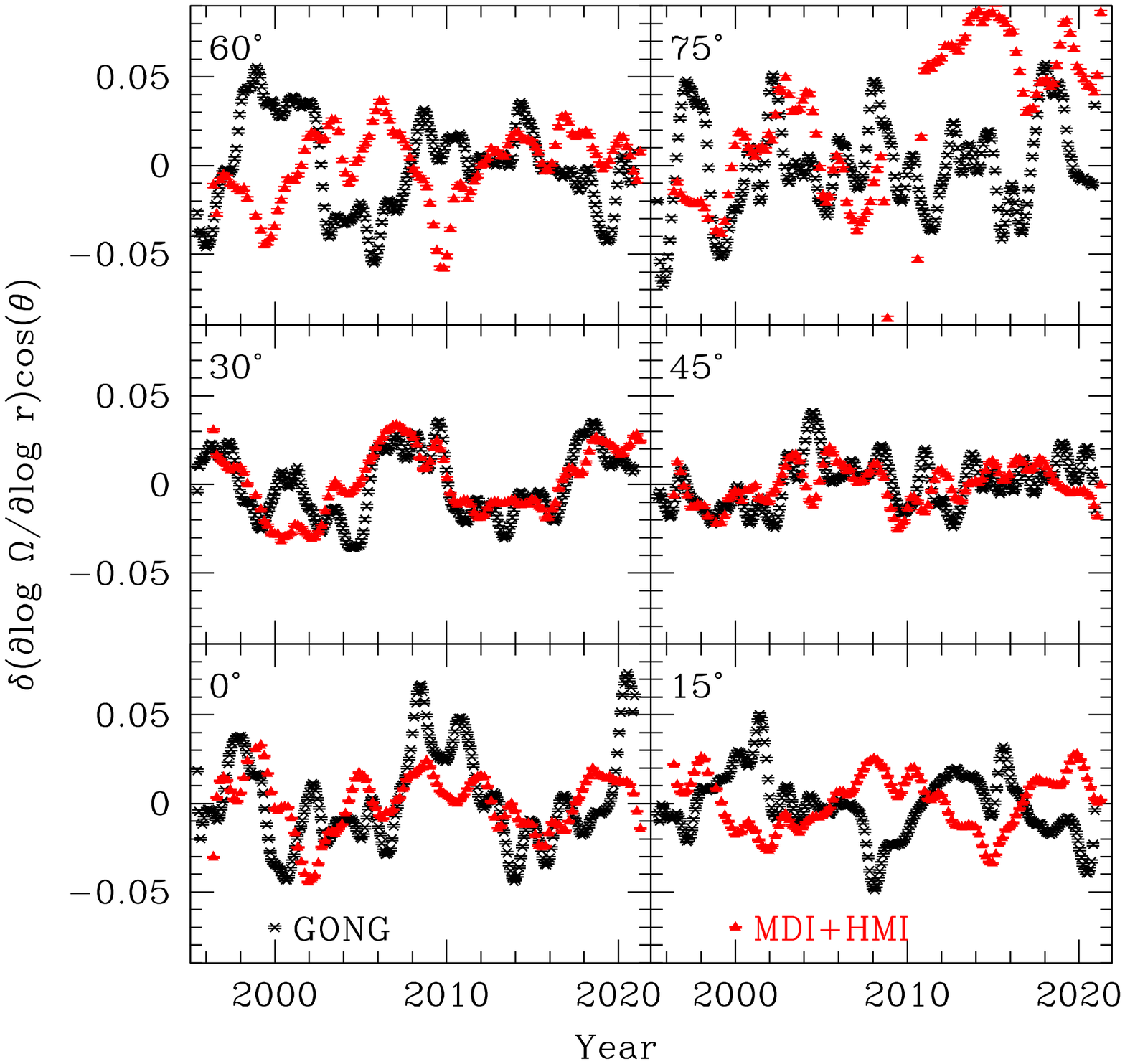}
\caption{The change in the dimensionless gradient at 0.97\rsun\ plotted as a function of time at a few different latitudes. The black points are results obtained with GONG data, while the red show  MDI and HMI results.
}
\label{fig:latcol}
\end{figure}

\begin{figure}
\epsscale{1.1}
\plotone{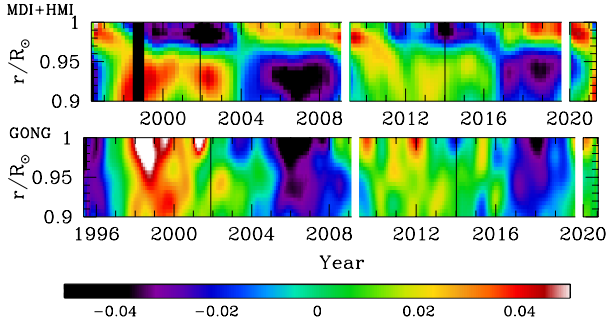}
\caption{The same as Fig.~\ref{fig:zon99}, now the results are shown for a latitude of $20^\circ$ as a function of time and radius. The upper panel is MDI+HMI, the lower panel is 
GONG. The lack of resolution of GONG inversions very close to the surface accounts for the very large differences there. The upper panel clearly shows how the gradient changes within the NSSL. 
}
\label{fig:zon20}
\end{figure}

\begin{figure}
\epsscale{1.1}
\plotone{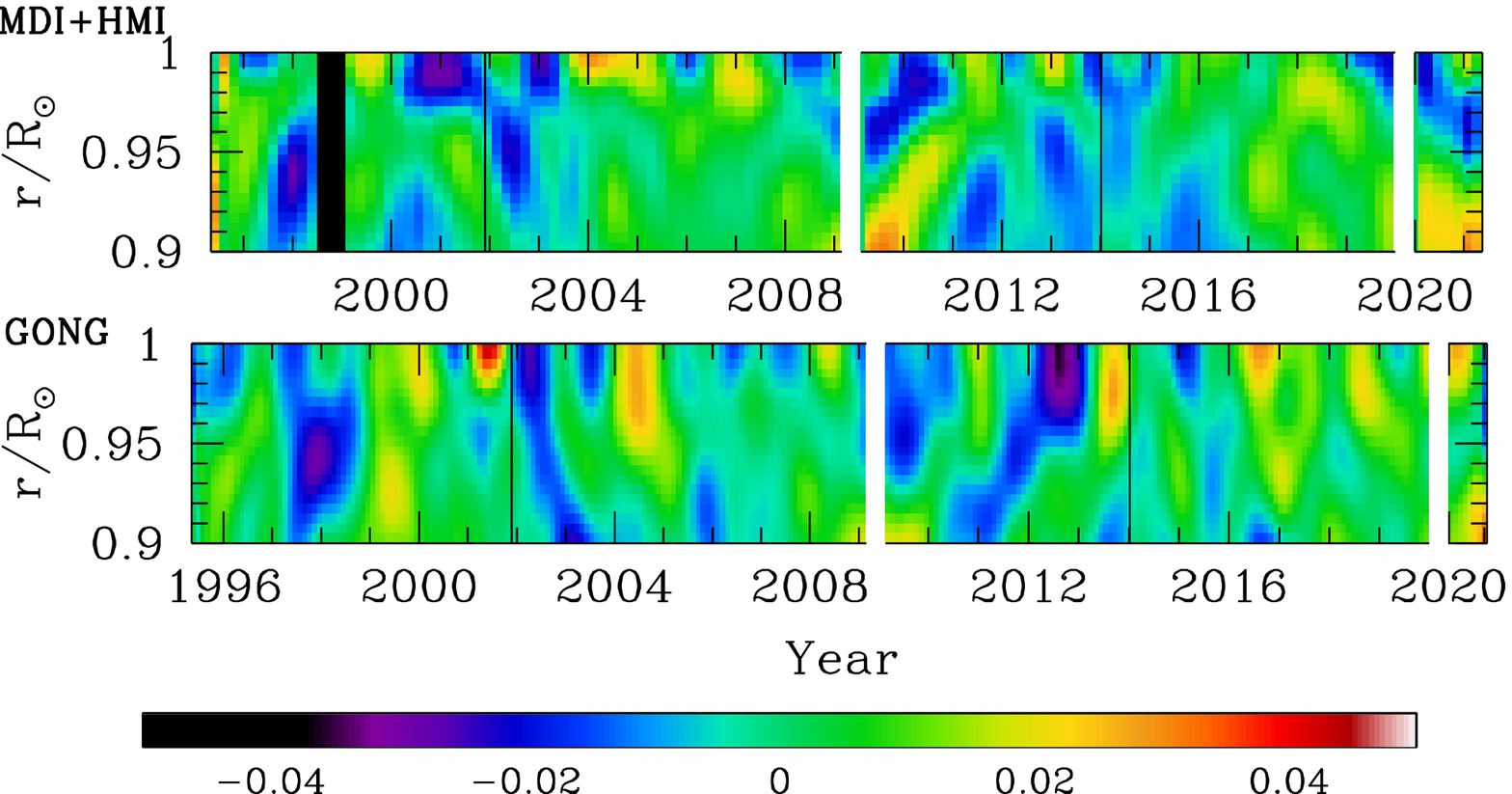}
\caption{The same as Fig.~\ref{fig:zon20}, but for $40^\circ$ latitude.
}
\label{fig:zon40}
\end{figure}

In Fig.~\ref{fig:latcol} we show the change of gradient (again multiplied by $\cos\theta$ for consistency with the other figures) at 0.97\rsun\ at a few selected latitudes. The figure clearly shows quasi-periodic changes at low latitudes. Although the high-latitude results are dominated by systematic errors in the data sets, there appears to be a hint of periodic behavior even at these latitudes. What we do not completely understand is the disagreement between GONG and MDI+HMI results at $15^\circ$; it is possible that this is because GONG results have poor resolution at this depth --- a similar figure plotted at 0.95\rsun\ does not show such a disagreement at low- and mid-latitudes. However, another reason for the disagreement is probably the slight oscillations in the GONG average gradients (Fig:~\ref{fig:ave}) that feed into the results of the change in the gradient. These oscillations at $0.97R_\odot$ are a result of using only 8 splitting coefficients to obtain results with GONG data, compared with 18 coefficients for MDI/HMI. Thus, the MDI and HMI results may be better in terms of examining the time variation. 

Figures~\ref{fig:zon99}\ and \ref{fig:zon95}\ imply that there is a depth dependence in the change of the gradient, this is shown in Figs.~\ref{fig:zon20} and \ref{fig:zon40} for latitudes of $20^\circ$ and $40^\circ$ respectively. At $20^\circ$ we see a change in the sign of the residual as one goes deeper in radius.

\begin{figure}
\epsscale{1.1}
\plotone{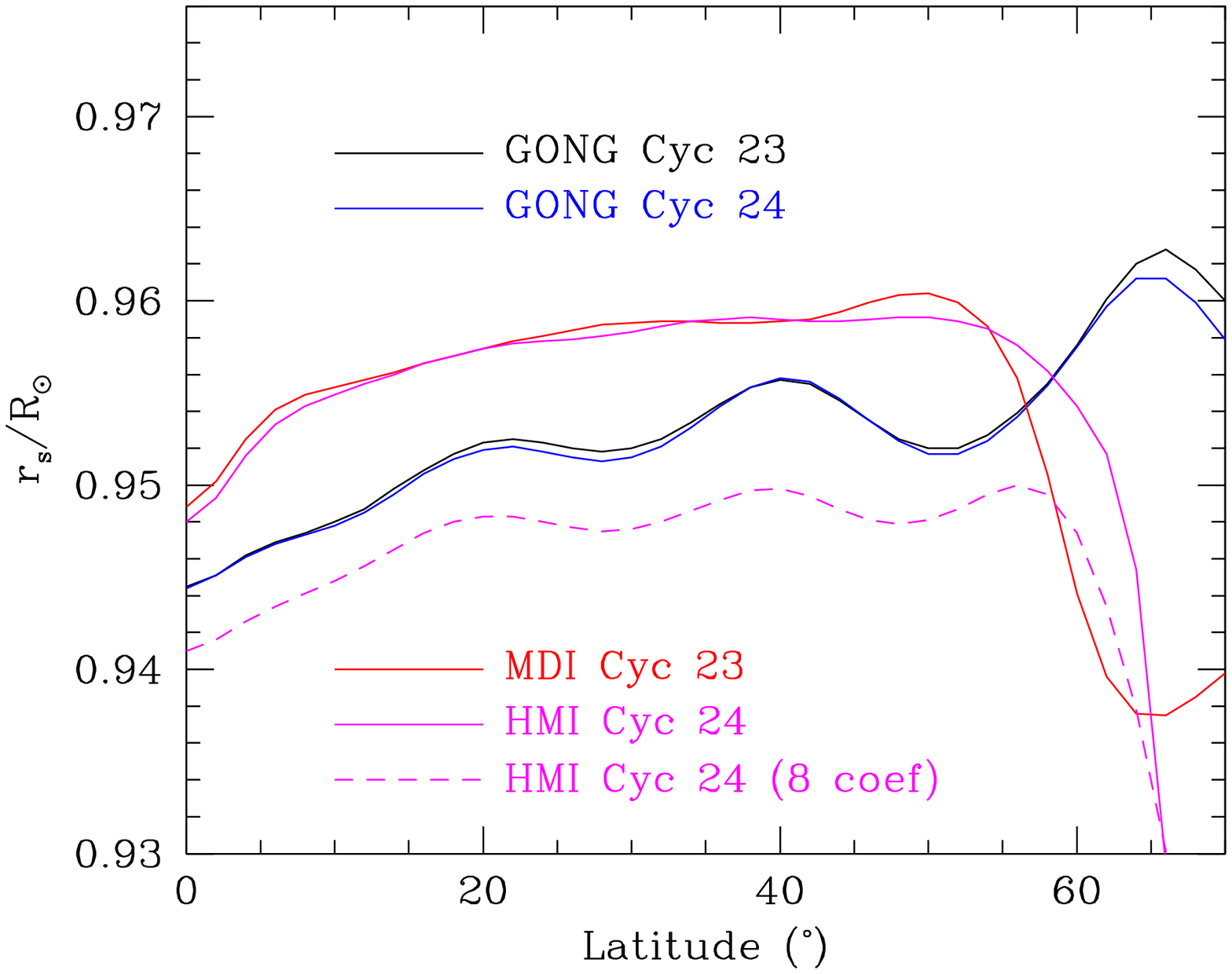}
\caption{The average radius of the base of the shear layer plotted as a function of latitude. We also show the result of restricting the  HMI data sets  to modes with $l \le 150$, and 8 splitting coefficients. 
}
\label{fig:rs}
\end{figure}

\begin{figure}
\epsscale{1.1}
\plotone{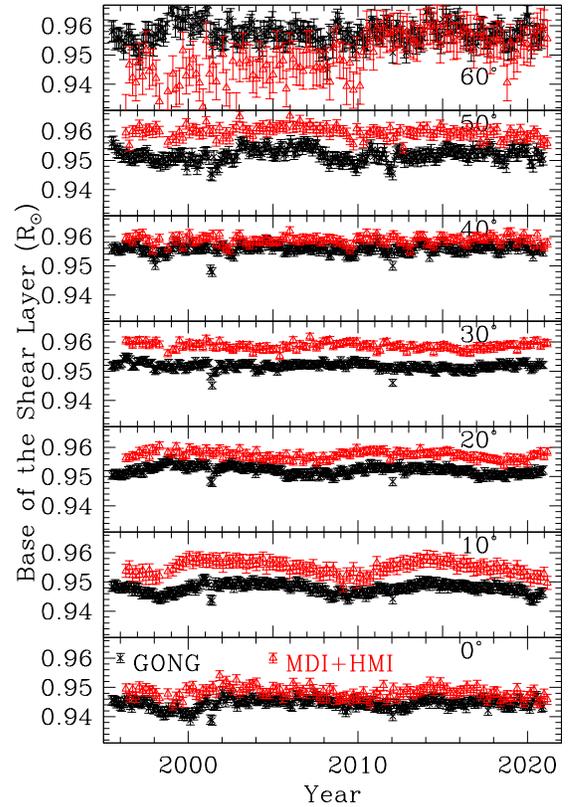}
\caption{The position of the base of the shear layer at different latitudes plotted as a function of time. The black crosses are results obtained with GONG data, the red triangles are for MDI and HMI data. 
}
\label{fig:depth}
\end{figure}

\begin{figure*}
\epsscale{0.75}
\plotone{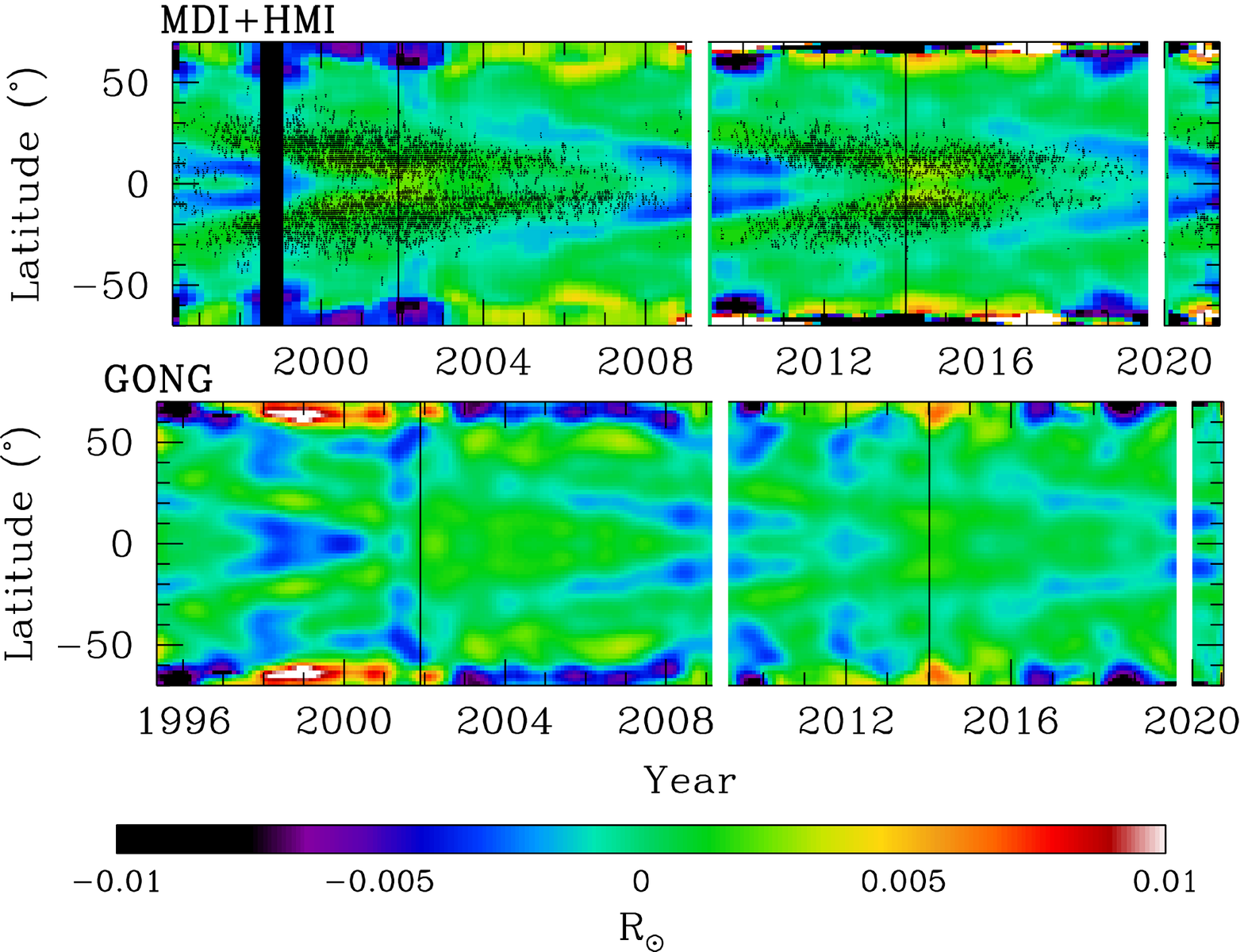}
\caption{The change in the extent of the near-surface shear layer as a function of time and latitude for MDI+HMI (upper panel) and GONG (lower panel). The figure shows the residuals obtained after the average for each cycle is subtracted. The vertical lines are the same as in Fig.~\ref{fig:zon99}. Note that the changes mimic the pattern of changes in Fig.~\ref{fig:zon99}, but are less well-defined.
}
\label{fig:time}
\end{figure*}

\subsection{The extent of the NSSL}
\label{subsec:depth}

The depth of the shear layer will depend on the physical conditions inside the convection zone, and may show variations when the conditions change, which happens with increasing magnetic activity. Hence, a study of the extent of NSSL may give information about the conditions in the near-surface layers of the Sun and provide constraints on theoretical models of NSSL.
The extent of the shear layer has a latitude dependence, and this can be seen in Fig.~\ref{fig:rs}. The shear layer is deepest at the equator and becomes shallower with increasing latitude. We also find systematic differences between results obtained with GONG data and those obtained with MDI and HMI data. The differences between GONG and MDI have been found earlier \citep[see e.g.,][]{hmasb2010}; there are also differences between GONG and HMI data, as well as MDI and HMI data \citep{sbhma2019}. One possible reason why
GONG data give a deeper estimate of the NSSL is GONG datasets  have very few  modes with lower-turning points shallower than $0.98R_\odot$ and consequently,  the surface value is essentially determined by extrapolation due to the regularization (smoothing) used in the RLS technique. As a result, the surface value of the gradient, which is used in estimating the depth of the shear layer, may not be correctly determined. The GONG results also show oscillations in latitude. These are caused because GONG sets are limited to 8 splitting coefficients; restricting the HMI sets to 8 splitting coefficients causes the
HMI result to be oscillatory, and this can also be seen in Fig.~\ref{fig:rs}. In addition to introducing oscillations as a function of latitude, restricting the mode set and the number of coefficients makes the inferred depth of the shear layer larger. There is also a discrepancy between the MDI and HMI results at high latitude, which produces a shift in $r_s$ above $60^\circ$ latitude; this is most likely to be a result of the known systematic differences between MDI and HMI splitting coefficients \citep[see  Fig.~1 of ][]{sbhma2019}.

The most well-defined solar-cycle related changes in the depth of NSSL are confined to low latitudes. The position of the bottom of the NSSL, $r_s$,  shows a change in the active latitudes, as can be seen in Fig.~\ref{fig:depth}; the changes are small just above the active latitudes, but appear to increase again at the highest latitudes. The nature of the shear layer appears to be complicated at high latitudes, though the reliability of the results here is questionable. \citet{hmasb2010} also found a time-variation in $r_s$, but with data available for only one solar cycle and with the numerical differentiation used resulting in larger uncertainties, the changes were not clear. The time variation also shows discrepancies at high latitudes.  

The change in $r_s$ at different latitudes is clearer if we subtract the time-average of $r_s$ and plot the residuals, i.e., $r_s-<r_s>$. The results are shown in Fig.~\ref{fig:time}. As can be seen, the changes in $r_s$ also show a migrating pattern, that is similar to, but not as well-defined, as the pattern of the changes in the gradient. The appearance of sunspots coincides with latitudes where the NSSL become shallower than average, i.e., where $r_s$ increases.

\section{Discussion}
\label{sec:disc}

Our investigation of the NSSL layer using both f- and p-modes of solar oscillations show that the gradient of solar rotation is mostly independent of latitude, but depends on radius --- the absolute value of the gradient decreases with radius. This is not surprising since the gradient is small below the NSSL, i.e., below about $0.95R_\odot$.

At low latitudes, our results agree with earlier helioseismic analyses, i.e., we find a logarithmic gradient of $-1$. The behavior of the NSSL at high latitudes is uncertain --- different data sets give different results (see Fig.~\ref{fig:ave}). Results obtained with GONG, MDI and HMI differ for latitudes higher than about $60^\circ$. \citet{barekat2} had examined systematic differences between MDI and HMI results and had concluded that the 72-day high-degree HMI data suffer from systematic errors. However, we find that the HMI and GONG results are more similar --- this is what \citet{sbhma2019} also found as seen in Fig.~1 of their paper. 

It should be noted that the behavior of the \citet{mjt}, \citet{barekat1} and \citep{barekat2} results are very different from ours at high latitude; their results, obtained with only f modes, show a decrease in the absolute value of the gradient at high latitudes, our results show the gradient becoming steeper around $60^\circ$.  It is likely that the inclusion of p-mode data changes the inferred high-latitude behavior of the gradient in the NSSL, and that there may be systematic differences between f- and p-mode data. There could also be systematic differences between the results obtained using p-modes and f-modes because f-modes of high degree are confined very close to the solar surface, where the reference model used to calculate the kernels may not be good enough. The difference between our high-latitude results and those of \citet{mjt} and \citet{barekat1,barekat2} implies that the results cannot be trusted. It may be possible to determine the gradient at the shallow depths at high latitudes with local helioseismic data from off-ecliptic missions like Solar Orbiter or the proposed polar mission, {Solaris\footnote{https://www.nasa.gov/press-release/nasa-selects-proposals-for-new-space-environment-missions}\citep{hassler2020, hassler2021} which aims to observe the solar polar regions for 3 months.}

The gradient in the NSSL changes with the solar cycle. The pattern of change is reminiscent of the zonal flow pattern, but with one major difference --- in the near-surface zonal flow pattern, the low-latitude bands migrate towards the equator, while the high-latitude one move towards the poles, however, the change of gradient in the NSSL shows both low-latitude and high-latitude bands moving towards the equator. This was also seen by \citet{barekat2}. This is clearer in the deeper layers of the NSSL; in the zonal flows, this feature is seen much deeper, at about 0.8\rsun\ \citep{sbhma2019}. We also find that not just the gradient of the NSSL, but the extent of the NSSL also changes with the solar cycle, and here too, the change shows a migrating pattern with time, with bands moving towards the equator.

Recreating the NSSL correctly in simulations of solar rotation has been challenging. Global MHD simulations are usually done under the anelastic approximation which does not allow simulating rotation and convection correctly in the outer few per cent of the Sun \citep[see, e.g.,][]{miesch}, missing out a large part of the NSSL. Such simulations show an NSSL if the density contrast is increased, but even then a shear layer is obtained only at low latitudes \citep{matilsky2019}. Fully compressible radiative-hydrodynamic simulations in sections of spherical shells are able to form an NSSL {close to the surface, but again, only} at low latitudes \citep{robinson}. \citet{rudiger}  find that they can form an NSSL using a quasi-linear theory. However, getting the correct magnitude of the gradient at different latitudes has not yet been possible, even with massively parallel, fully compressible codes.
There has been some work done to examine the effects of magnetic fields on the NSSL, notably \citet{kit} found that the shear in this layer is sensitive to the toroidal magnetic fields in these layers, and can serve as a probe for fields of the order of one Kilo Gauss. Our results should, therefore,  give constraints on the sub-surface magnetic fields. The change in the extent of the NSSL with activity is also a constraint. The presence of the NSSL has been explained as being due to Reynold stresses \citep{robinson} and the change in the gradient because of the suppression of turbulent viscosity by the magnetic field \citep{kit}, thus the suppression of turbulence by magnetic fields could potentially explain changes in the NSSL, but models of the NSSL need to explain both the change in the gradient and the change in the extent of the NSSL.

\section*{Acknowledgments}
{We thank the anonymous referee for comments that have helped to improve the paper.}
We would like to thank Richard S. Bogart for access to his easily usable archive of NOAA active region characteristics. This work utilizes data from the National Solar Observatory Integrated Synoptic Program, which is operated by the Association of Universities for Research in Astronomy, under a cooperative agreement with the National Science Foundation and with additional financial support from the National Oceanic and Atmospheric Administration, the National Aeronautics and Space Administration, and the United States Air Force. The GONG network of instruments is hosted by the Big Bear Solar Observatory, High Altitude Observatory, Learmonth Solar Observatory, Udaipur Solar Observatory, Instituto de Astrofísica de Canarias, and Cerro Tololo Interamerican Observatory. This work also uses data from the Michelson Doppler Imager on board SOHO. SOHO is a project of international cooperation between ESA and NASA. {We also use data from the Helioseismic and Magnetic Imager on board the Solar Dynamics Observatory. HMI data are courtesy of NASA/SDO and the AIA, EVE, and HMI science teams.}

\facility{GONG, MDI, HMI}

\bibliography{main}{}
\bibliographystyle{aasjournal}

\end{document}